___

# SPEAKER RECOGNITION USING RESIDUAL SIGNAL OF LINEAR AND NONLINEAR PREDICTION MODELS[1]

*Marcos Faundez-Zanuy, Daniel Rodríguez-Porcheron*

Escola Universitària Politècnica de Mataró
Universitat Politècnica de Catalunya
MATARO (BARCELONA) SPAIN
faundez@eupmt.es  http://www.eupmt.es/veu

## ABSTRACT

This Paper discusses the usefulness of the residual signal for speaker recognition. It is shown that the combination of both a measure defined over LPCC coefficients and a measure defined over the energy of the residual signal gives rise to an improvement over the classical method which considers only the LPCC coefficients. If the residual signal is obtained from a linear prediction analysis, the improvement is 2.63% (error rate drops from 6.31% to 3.68%) and if it is computed through a nonlinear predictive neural nets based model, the improvement is 3.68%.

## 1. INTRODUCTION

Although the relevance of the LPC residual signal for speech coding (Multi-pulse LPC, CELP, etc.) is well established, little attention has been dedicated to this signal for speaker recognition purposes.

In [1] it has been shown that humans can recognize people by listening to the LPC residual signal, and several authors have used pitch frequency for speaker recognition. On the other hand, it has been found [2] that the residue as a whole carries richer information than the fundamental frequency alone, and the use of a cepstrum computed over the LPC residual signal has been proposed. The use of this parameterization is less efficient than the cepstrum of the LPC coefficients, but a combination of LPC cepstrum and LPC residual cepstrum produces a reduction in the error rate from 5.7% to 4.0%. Although in [2] the residual signal is used, the residue of the LPC residual cepstrum is ignored. For this reason we propose:

a) The use of a measure error defined over the LPC-residual signal, (instead of a parameterization over this signal) combined with a classical measure defined over LPCC coefficients. We have found that these two kinds of measures are uncorrelated and complementary. Their combination reduces the error rate in 1%. Our system is based on a vector quantizer approach. That is, each speaker is modeled with one codebook.

b) The use of a nonlinear prediction model based on neural nets, which has been successfully applied to a waveform speech coder [3]. It is well known that the LPC model is unable to describe the nonlinearities present in the speech, so useful information is lost with the LPC model alone. With a nonlinear prediction model based on neural nets it is not possible to compare the weights of the neural net. This is due to the fact that infinite sets of different weights representing the same model exist, and direct comparison is not feasible. For this reason the measure is defined over the residual signal of the nonlinear prediction model. For improving performance upon classical methods a combination with linear parameterization must be used. In order to reduce the computational complexity and to improve the recognition rates a novel scheme, which consists of the preselection of the K speakers nearest to the test sentence is proposed. Then, the error measure based on the nonlinear predictive model is computed only with these speakers. ( In this case a reduction of 3.68% in error rate upon classical LPC cepstrum parameterization is achieved).

## 2. SPEAKER RECOGNITION USING LPCC COEFFICIENTS AND RESIDUAL SIGNAL

### 2.1 Database

Our experiments have been computed over 38 speakers from the New England dialect of the DARPA TIMIT Database (24 males&14 females). The speech samples were downsampled from 16KHz to 8 Khz, and pre-emphasized by a first order filter whose transfer function was $H(z)=1-0.95z^{-1}$. A 30ms Hamming window was used, and the overlapping between adjacent frames was 2/3. A cepstral vector of order 12 was computed from the LPC coefficients. Five sentences are used for training, and 5 sentences for testing (each sentence is between 0.9 and 2.8 seconds long).

### 2.2 Recognition algorithm

Our recognition algorithm is a Vector Quantization approach. That is, each speaker is modeled with a codebook in the training process. During the test, the input sentence is quantified with all the codebooks, and the codebook which yields the minimal accumulated error indicates the recognized speaker.

The codebooks are generated with the splitting algorithm. Two methods have been tested for splitting the centroids:
a) The standard deviation of the vectors assigned to each cluster.
b) A hyperplane computed with the covariance matrix (See [4])

Table 1 compares the error rates with both methods, and the results reported by Farrell in [5]. Our results compare favorably with Farrell ones, because we use the 38 speakers of the New England

___

[1] This work has been supported by the CICYT TIC97-1001-C02-02

___

dialect whereas he uses only 20. In what follows we will consider the Hyperplane method for splitting the centroids.

| codebook size | standard deviation | Hyperplane | Farrell |
|---|---|---|---|
| 4 | 10,5% | 10% | 10% |
| 5 | 6,31% | 6,84% | 8% |
| 6 | 6,31% | 6,31% | 5% |
| 7 | 6,84% | 3,68% | 4% |

**Table 1:** Identification errors (%)

The next step is the evaluation of the performance with different error criteria. We have tested the following:

*Measures defined over the coefficients*
measure 1 (1): Mean Square Error (MSE) of the LPCC.
measure 2 (2): Mean Absolute difference (MAD) of the LPCC.

*Measures defined over the residue*
measure 3 (3): MSE of the residue.
measure 4 (4): MAD of the residue.
measure 5 (5): Maximum absolute value (MAV) of the residue.
measure 6 (6): Variance ($\sigma$) of the residue.
Where the residue is obtained by filtering the input frame with the LPC coefficients.
Table 2 summarizes the results for the different measures.

| codebook size U | 1 | 2 | 3 | 4 | 5 | 6 |
|---|---|---|---|---|---|---|
| 4 | 10 | 7,89 | 42,10 | 23,68 | 76,84 | 42,10 |
| 5 | 6,84 | 5,26 | 37,36 | 21,05 | 77,89 | 37,36 |
| 6 | 6,31 | 4,21 | 32,10 | 20,53 | 74,74 | 32,10 |
| 7 | 3,68 | 3,16 | 26,32 | 14,21 | 82,10 | 26,32 |

**Table 2:** Identification errors (%) as function of the measure and codebook size

Main conclusions:

- Measures defined over the residual signal are less efficient than measures defined over the LPCC. In figure 1 it can be seen that inter and intra-speaker distortions are very similar for measures defined over the residue, whereby is difficult to distinguish between speakers using only a measure defined over such a residue.

- Speech signal has zero mean, so the variance is equivalent to the MSE (measures 3 and 6).

- Alltough the measures defined over the residue give lower recognition rates, they are low correlated with measures defined over the lpcc coefficients (see table 3 and dispersion diagram of figure 2) and therefore they can be combined in order to improve the behaviour of the system. It is important to see that there is a low correlation between measures defined over different signals (residue and coefficients), and high correlation when different measures are defined over the same information (coefficients or residue). This entails that it is more interesting to combine measures defined over the residue and the coefficients than to use different measures defined over the same signal.

| $\rho$ | 2 | 3 | 4 | 5 | 6 |
|---|---|---|---|---|---|
| 1 | 0.98 | 0,14 | 0,23 | 0,09 | 0,14 |
| 2 |  | 0.15 | 0,23 | 0,09 | 0,15 |
| 3 |  |  | 0,91 | 0,87 | 1 |
| 4 |  |  |  | 0,95 | 0,91 |
| 5 |  |  |  |  | 0,88 |

**Table 3:** correlation coefficients between measures por 7 bit codebooks.

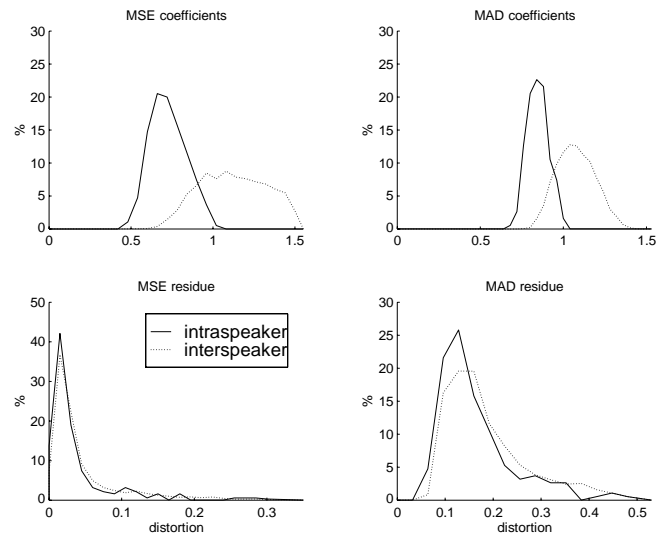

**Figure 1:** Histograms of inter and intra-speaker for codebooks of 7 bits.

| Number of bits | measures (2,3) | measures (2,4) |
|---|---|---|
| 4 | 7,89 | 7,89 |
| 5 | 4,21 | 4,74 |
| 6 | 3,68 | 3,68 |
| 7 | 3,16 | 3,16 |

**Table 4:** error rates (%) as function of codebook size, for 2 different combinations.

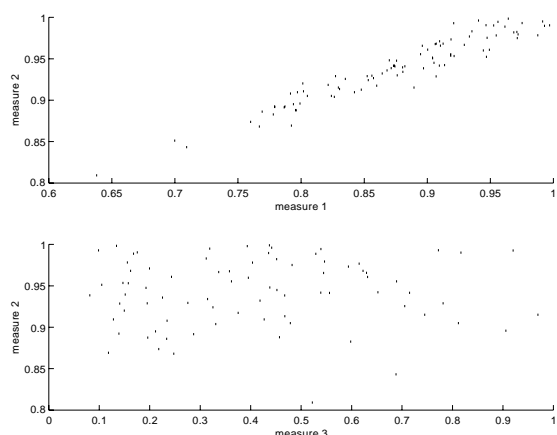

**Figure 2:** dispersion diagrams for combinations of different measures for 7 bits codebook

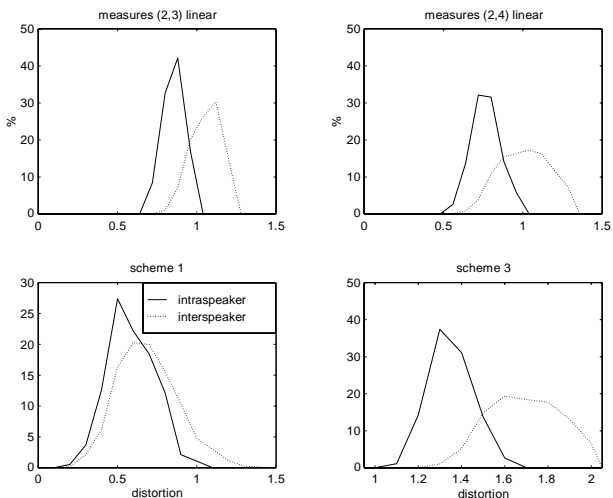

**Figure 3:** Histograms

Table 4 presents the results obtained by combining the best measures over coefficients and residue. An important result is the reduction of 2.63% in the recognition error for 5 and 6 bit codebooks (combination (2,3)).

## 3. SPEAKER RECOGNITON USING LPCC COEFFICIENTS AND NONLINEAR PREDICTION RESIDUAL SIGNAL

In [6] we studied the possibility of modeling each speech frame with a nonlinear predictive model based on a multilayer perceptron neural net. It obtained a significative improvement in prediction gain upon the classical LPC analysis. In [3] we applied this model to an ADPCM speech waveform coder, changing the linear predictor to a nonlinear predictor. The results outperform the linear prediction between 1 and 2 dB in SEGSNR. In [7] we present an efficient algorithm for reducing the computational complexity of computing the nonlinear predictive model and in [8] we discuss a sample adaptive ADPCM scheme. Here, we will apply the nonlinear predictive model to a speaker recognition problem, based on vector quantization.

### 3.1 Nonlinear codebook generation

In order to generate the codebook a good initialization must be achieved. Thus, it is very important to achieve a good clustering of the train vectors. We have evaluated several possibilities, and the best one is to implement first a linear LPCC codebook of the same size. This codebook is used for clustering the input frames, and then each cluster is the training set for a multilayer perceptron with 10 input neurons, 4 neurons in the first hidden layer, 2 neurons in the secod hidden layer, and one output neuron with a linear transfer function. Thus, the MLP is trained in the same way as in [3], but the frames have been clustered previously with a linear LPCC codebook. After this process, the codebook can be improved with a generalization of the Lloyd iteration:
a) frames are clustered with the nonlinear codebook.
b) new neural nets are trained with the new clusters.

In this process, the weights and biases of the network are initialized with a multi-start algorithm, which consists in training 4 different random initializations and the weights of the previous iteration (5 different initializations). After the training process, we choose the initialization that gives the lowest quantization distortion. For each initialization 8 epochs are done with the Levenberg-Marquardt algorithm.
With this codebook several schemes have been tested.

*Scheme 1*
The nonlinear codebook is directly applied computing one codebook for each speaker during the training process. In the test phase, the input sentence is quantized with each codebook. The codebook which yields the lowest accumulated error is selected. Mean absolute difference of the residual signal of the nonlinear prediction process is the result of filtering the frames with each vector of the codebook (neural net) and choosing the lowest MAD. Table 5 summarizes the results for different codebook sizes, applying the generalized Lloyd iteration for improving the codebooks. The results of table 5 are not comparable to the LPCC codebook of table 1, so new schemes have been tested.

| Iteration 0 | | | Iteration 1 | | | Iteration 2 | | | Iteration 3 | | |
|---|---|---|---|---|---|---|---|---|---|---|---|
| 4 | 5 | 6 | 4 | 5 | 6 | 4 | 5 | 6 | 4 | 5 | 6 |
| 24 | 33 | 28 | 24 | 25 | 22 | 24 | 24 | 23 | 23 | 24 | 24 |

**table 5:** Identification errors (%) for different codebook sizes (between 4 and 6 bits) for different iterations.

*Scheme 2*
The LPCC used for clustering the frames is used as pre-selector of the recognized speaker. That is, the input sentence is quantized with the LPCC codebooks and the K codebooks that produce the lowest accumulated error are selected. Then, the input sentence is quantized with the K nonlinear codebooks, and the accumulated distance of the nonlinear codebook is selected as the error criterion. With this system, for K=2 the results are (table 6):

| Linear | Iteration 0 | | | Iteration 3 | | |
|---|---|---|---|---|---|---|
| | MLP Codebook | | | MLP Codebook | | |
| | 4 | 5 | 6 | 4 | 5 | 6 |
| 4 bits | 15.79 | 21.58 | 16.84 | 17.37 | 17.89 | 16.84 |
| 5 bits | 14.74 | 21.58 | 17.89 | 14.21 | 15.79 | 15.79 |
| 6 bits | 14.74 | 22.63 | 17.89 | 14.74 | 15.26 | 15.26 |
| 7 bits | 15.79 | 23.16 | 18.95 | 14.74 | 16.84 | 16.84 |

**table 6:** Identification errors (%) for different codebook sizes (between 4 and 6 bits) and iterations 0 and 3.

The computational complexity of filtering each frame with the neural nets is higher than the comparison of the LPCC coefficients. Thus scheme 2 implies a reduction of the required number of flops, because the input frames are not filtered with all the codebooks like in scheme 1

_________________________________________________________________________________________

These results are not good enough compared with the results of table 1. For this reason we decided to implement an hybrid structure, combining an error measure defined over the LPCC coefficients and an error meause defined over the residue of the nonlinear predictive analysis.

*Scheme 3*

This scheme is similar to scheme 2, but thedistortion information obtained from the LPCC coefficients is combined with the error measure defined over the residue, with the following expression:

$error = LPCC\_error + \alpha * residue\_error$ ,where the combination factor has been determined experimentally, by a trial and error procedure, and its value has no special meaning because the combined terms have different origins: the LPCC errors are obtained computing the difference between vectors of dimension 12, and each component has a small value. The residue error is obtained over 240 samples, and each sample have greater magnitudes than LPCC coefficients.

table 7 summarizes the obtained results with K=2.

| Linear Codebook | Iteration 0 MLP Codebook | | | Iteration 3 MLP Codebook | | |
|---|---|---|---|---|---|---|
| | 4 | 5 | 6 | 4 | 5 | 6 |
| 4 bits | 7.89 | 7.37 | 7.89 | 7.37 | 6.84 | 7.89 |
| 5 bits | 3.68 | 3.16 | 4.74 | 3.68 | 4.21 | 4.74 |
| 6 bits | 3.68 | 3.16 | 4.21 | 3.68 | 3.68 | 3.68 |
| 7 bits | 3.16 | 2.63 | 2.63 | 3.16 | 2.63 | 2.63 |

**Table 7:** Identification errors (%) for different codebook sizes (between 4 and 6 bits) and iterations 0 and 3, for scheme 3.

This scheme outperforms the scheme based on LPCC of table 1 and the combined scheme with linear residue of table 4. The most relevant conclusions are:

- There is a reduction in error rate of 1% compared with the LPCC codebook of 7 bits, and 3% with respect to 4, 5 and 6 bits LPCC codebooks.

- There is a reduction in error rate of 0.5% compared with the best combination between linear residue and LPCC coefficients, for a codebook of 7 bits. For codebooks of 4, 5 and 6 bits the improvement is about 1%.

Figure 3 shows the histograms of interspeaker and intraspeaker distortions. On the top for a combination of residue's MAD and LPCC distortion. On the bottom for schemes 1 and 3 of nonlinear predictive models.

It is interesting to see that the use of the residue in the nonlinear predictive model implies the use of the residual information, and also of the nonliear predictive model, because it affects the magnitude of the residue.

## 4. CONCLUSIONS

In this paper we have evaluated the relevance of the residual signal of LPC analysis for speaker recognition purposes, and the hability of nonlinear predictive models for speaker recognition. We believe that the structure can be simplified if the nonlinear codebook generation process is improved. Whereby it is important to obtain a good clustering algorithm, evaluate the influence of the number of epochs, network architecture, etc. This paper, is only a first approach of the nonlinear predictive models to speaker recognition.